\documentclass[12pt]{article}
\usepackage{amsmath}
\usepackage{amssymb}
\begin{document}
\begin{center}
{\bf {\large{A 3D Field-Theoretic Example for Hodge Theory}}} 

\vskip 3 cm

{\sf  A. K. Rao$^{(a)}$,  R. P. Malik$^{(a,b)}$}\\
$^{(a)}$ {\it Department of Physics, Institute of Science,}\\
{\it Banaras Hindu University (BHU), Varanasi - 221 005, (U.P.), India}\\

\vskip 0.1cm

$^{(b)}$ {\it DST Centre for Interdisciplinary Mathematical Sciences,}\\
{\it Institute of Science, Banaras Hindu University (BHU), Varanasi - 221 005, India}\\
{\small {\sf {e-mails:  amit.akrao@gmail.com;  rpmalik1995@gmail.com}}}
\end{center}

\vskip 2.0 cm

\noindent
{\bf Abstract:}
We focus on the continuous  symmetry transformations for the three ($2 + 1$)-dimensional 
(3D) system of a combination of the {\it free} Abelian 1-form and 
2-form gauge theories within the framework of Becchi-Rouet-Stora-Tyutin (BRST) formalism.  
We establish that this {\it combined} system  is a tractable field-theoretic model of Hodge theory. The symmetry operators 
of our present system provide the 
physical realizations of the de Rham cohomological operators of differential geometry at the algebraic level. Our present
investigation is important in the sense that, for the first time, we are able to establish an {\it odd} dimensional 
(i.e. $D = 3$) field-theoretic
system to  be an example for Hodge theory (besides  earlier works on a few interesting  ($0 + 1$)-dimensional (1D) toy models as well as 
a set of well-known ${\mathcal N}  = 2$ SUSY quantum mechanical systems of physical interest). For the sake of brevity,  we have purposely {\it not} taken into account the 3D Chern-Simon term for the Abelian 1-form gauge field in our theory which allows the mass as well as the gauge-invariance to co-exist together.

\vskip 1 cm
\noindent
PACS numbers:  03.70.+k, 11.30.-j, 02.40.-k

\vskip 1 cm
\noindent
{\it {Keywords}}:  3D Abelian 1-form and  2-form gauge theories;  off-shell
nilpotent (anti-)BRST symmetries; off-shell nilpotent (anti-)co-BRST symmetries; a unique bosonic symmetry; ghost-scale symmetry;  
cohomological operators; algebraic structures

\newpage

\section{Introduction}
The (super)string theories (see, e.g. [1-3] and references therein) are the forefront areas of research activities in the 
modern day theoretical high energy physics (THEP) which have brought together researchers in the domains of THEP
and mathematics on a single intellectual platform. The modern developments in these theories have led to many other areas of 
research activities in THEP and pure mathematics. One of them is the study of the higher $p$-form ($p = 2, 3,...$) gauge theories because
the higher $p$-form basic fields appear in the quantum excitations of (super)strings. Hence, very naturally, the reach
and range of the (super)string theories go beyond the realm of the standard model of particle physics which is
based on the interacting non-Abelian 1-form (i.e. $p = 1$) gauge theory. In this connection, it is
pertinent to point out that,  during the last few years, we have devoted time 
on the research activities that are connected with the Becchi-Rouet-Stora-Tyutin (BRST) approach to the higher $p$-form 
($p = 2, 3, ...$) gauge theories and established that the massless and St$\ddot u$ckelberg-modified massive 
Abelian $p$-form ($p = 1, 2, 3, ...$) gauge theories in $D = 2 p$ dimensions of spacetime are the 
field-theoretic examples for Hodge theory where there is a confluence  of ideas from the physics part
of the BRST formalism [4-7] and mathematical aspects of differential geometry at the algebraic level. 
Hence, we observe that, in our studies on the models of Hodge theory (see, e.g. [8-10] and reference therein), the ideas from 
physics and mathematics are beautifully intertwined together in a meaningful manner.

Against the backdrop of the above discussions, it is clear that the {\it above}
field-theoretic models of Hodge theory are defined only in the even (i.e. $D = 2p$) dimensions of spacetime.
The purpose of this Letter is to establish that the BRST-quantized 
{\it combined} system of the free Abelian 1-form and 2-form gauge theories is a tractable field-theoretic example for 
Hodge theory (within the framework of BRST formalism) in the three (2 + 1)-dimensional (3D) spacetime. 
This is for the first time, we are able to show an odd-dimensional ($D = 3$) field-theoretic system to 
be a model for Hodge theory (besides earlier works [11-13] on the one (0 + 1)-dimensional (1D) systems
of a rigid rotor [11], the celebrated FLPR model [12] and the dynamics of a particle on a torus [13]). These 1D systems
have also been shown to be the examples for Hodge theory. In addition, we have also devoted time on the proof  
that the ${\mathcal N}  = 2$ SUSY quantum mechanical systems {\it also} belong to this class of Hodge theory
(see, e.g. [14,15] and references therein). In our present investigation,  we show that our present 3D 
field-theoretical model provides the physical realizations of the de Rham cohomological operators\footnote{On a compact manifold without a boundary, 
a set of three operators ($d, \delta, \Delta$) belong to the de Rham cohomological operators which satisfy the algebra: $d^2 = \delta^2 = 0, \;
\Delta = (d + \delta)^2 = \{d, \, \delta |], \; [\Delta, \, d] = [\Delta, \, \delta] = 0$ where $(\delta)d$ are christened as the 
(co-)exterior derivatives and $\Delta$ is known as the Laplacian operator of differential geometry [16-18]. The (co-)exterior derivatives are
connected by the relationship: $ \delta = \pm\, *\, d\, *$ where $*$ is the Hodge duality operator (that is defined on the
above compact manifold without a boundary).} of 
differential geometry  (see, e.g. [16-18]) in terms of the continuous symmetry transformations 
(where there are total {\it six} of them). If the beauty of a theory is defined 
in terms of the numbers and varieties of 
symmetries it respects, the models of Hodge theory belong to this class.

Some of the key observations of our present endeavor are as follows. We note that the kinetic terms (owing their origin to the exterior derivative of differential geometry) for the Abelian 1-form and 2-form theories remain invariant under the (anti-)BRST  symmetry transformations. On the other hand,
the gauge-fixing terms (owing their origin to the co-exterior derivative of differential geometry) for the Abelian 1-form and 2-form gauge-fields
remain invariant under the (anti-)co-BRST symmetry transformations. The (anti-)ghost fields {\it either} do not transform 
at all {\it or} transform up to 
a U(1) gauge-type transformation under the {\it unique} bosonic symmetry transformation (which is nothing but an appropriate anticommutator of the nilpotent 
(anti-)BRST and (anti-)co-BRST symmetry transformations). The characteristic feature of the ghost-scale symmetry transformation is the observation
that {\it only} the fields of the ghost-sector transform under it (and fields from the non-ghost sector do {\it not} transform at all). The algebraic structures,
obeyed by these continuous symmetry operators, are reminiscent of the Hodge algebra that is respected by the cohomological operators of differential geometry.
Hence, our present 3D system (of the BRST-invariant free Abelian 1-form and 2-form gauge theories) provides a 
tractable field-theoretic example for Hodge theory.

The theoretical contents of our  present endeavor  are organized in the 
following order. First of all, in section two, we recapitulate the essentials of the (anti-)BRST 
symmetry transformations [$s_{(a)b}$] from our earlier work [19]. Our section three is devoted to the discussion on the 
(anti-)co-BRST symmetry transformations [$s_{(a)d}$]. The subject matter of our section four is connected 
with the derivation of a unique bosonic symmetry transformation $(s_w)$ from the appropriate anticommutators between the nilpotent 
(anti-)BRST and (anti-)co-BRST transformations. In section five, we devote time on the derivation of the
ghost-scale symmetry transformations (and their infinitesimal versions). Our section six deals with the algebraic structures 
of the symmetry operators of our theory and their connection with the Hodge algebra of cohomological operators.
Finally, in section seven, we summarize our key results 
and point out the future prospective of our present investigation.\\

\section{(Anti-)BRST Symmetries: Key Aspects}

We begin with the BRST-invariant Lagrangian density\footnote{For the flat 3D  Minkowskian background spacetime manifold, we choose the 
metric tensor $\eta_{\mu\nu} =  $ diag $(+ 1, - 1 , -1)$ so that the dot product between two non-null vectors 
$P_\mu$ and $Q_\mu$ is $P \cdot Q = \eta_{\mu\nu} \, P^{\mu}\, Q^{\nu}   \equiv P_0\,Q_0 - P_i \, Q_i $
where the Greek indices $\mu, \nu, \sigma, ... = 0, 1, 2$ stand for the time and space directions of our 3D spacetime manifold  and the 
Latin indices $i,  j, k = 1, 2$ correspond to space directions only. We have chosen the 3D Levi-Civita tensor 
$\varepsilon_{\mu\nu\sigma}$ such that $\varepsilon_{012} = + 1 = \varepsilon^{012}$ and it satisfies the standard 
relationships: $\varepsilon^{\mu\nu\sigma}\, \varepsilon_{\mu\nu\sigma} = 3 !, \; 
\varepsilon^{\mu\nu\sigma}\, \varepsilon_{\mu\nu\eta} = 2 ! \,\delta^\sigma_\eta, \;
\varepsilon^{\mu\nu\sigma}\, \varepsilon_{\mu\kappa\eta} = 1! \, (\delta^\nu_\kappa \, \delta^\sigma_\eta -
 \delta^\nu_\eta \, \delta^\sigma_\kappa)$ on the 3D flat Minkowskian  spacetime manifold (which remains 
in the background and does {\it not} participate in our discussions on symmetries).  } which is  the sum 
(i.e. ${\cal L}_{B} = {\cal L}^{(1)}_{B} + {\cal L}^{(2)}_{B} $) of the BRST-invariant
Lagrangian densities for the free Abelian 1-form and 2-form gauge theories as follows
\begin{eqnarray*}
{\cal L}_{B} &=& -\frac{1}{4} F^{\mu \nu} F_{\mu \nu} +  \frac{B^{2}}{2}   -  B (\partial \cdot A)
- \partial_{\mu} \bar{C} \, \partial^{\mu} C  \nonumber\\
&+& \mathcal{B} \left(\frac{1}{2} \varepsilon^{\mu \nu \sigma} \partial_{\mu} B_{\nu \sigma}\right)  
-\frac{\mathcal{B}^{2}}{2} 
+ B^{\mu} \left(\partial^{\nu} B_{\nu \mu}-\partial_{\mu} \phi \right) 
\nonumber\\
\end{eqnarray*}
\begin{eqnarray}
&-& \frac{B^{\mu} B_{\mu}}{2} 
+ \partial_{\mu} \,  \bar{\beta} \,\partial^{\mu} \beta +\left(\partial_{\mu} \bar{C}_{\nu} 
-\partial_{\nu} \bar{C}_{\mu}\right)\left(\partial^{\mu} C^{\nu}\right)  
\nonumber\\
&+& \left (\partial \cdot \bar{C}+\rho \right)\,  \lambda + \left (\partial \cdot C - \lambda \right )\, \rho, 
\end{eqnarray}
where ${\cal L}^{(1)}_{B} =  -\frac{1}{4} F^{\mu \nu} F_{\mu \nu} + \frac{1}{2}\, B^2  -  B (\partial \cdot A) 
- \partial_{\mu} \bar{C} \, \partial^{\mu} C$ is the (anti-)BRST invariant Lagrangian density for the free
Abelian 1-form  gauge theory and rest of the terms of the above Lagrangian density (1) are for the BRST-invariant
Lagrangian density in the case of a 3D free Abelian 2-form theory. As a side remark, we would like to mention that
the above {\it combined} system of the BRST-invariant Lagrangian density for the Abelian 1-form and 2-form gauge theories is 
the limiting case of the BRST invariant Lagrangian density that has been considered 
in our recent work [19] on the BRST approach to the St$\ddot u$ckelberg-modified 3D massive Abelian 2-form
theory with the conditions: $\tilde\varphi = 0,  \, m = 0$ and $\phi_\mu = A_\mu$ so that $\Sigma_{\mu\nu} = F_{\mu\nu}$. Here the field 
$\tilde\varphi$ corresponds to the pseudo-scalar field, $m$ denotes the rest mass of the Abelian 2-form field and the 
Lorentz vector field $\phi_\mu$ stands for the
St$\ddot u$ckelberg field (with its field-strength tensor as: $\Sigma_{\mu\nu} = \partial_\mu \phi_\nu - \partial_\nu \phi_\mu$). 
The special feature of the  3D Abelian 2-form theory is that the kinetic term for the antisymmetric ($B_{\mu\nu} = -\, B_{\nu\mu}$) tensor 
field $B_{\mu\nu}$ turns out to be the following 
\begin{eqnarray}
\frac{1}{12}\, H^{\mu\nu\lambda}\, H_{\mu\nu\lambda} = \frac{1}{2}\, H^{012}\, H_{012} = \frac{1}{2}\, (H_{012})^2,
\end{eqnarray}
where the covariant form of $H_{012} = \frac{1}{2}\, \varepsilon^{\mu\nu\sigma}\, \partial_\mu\,
B_{\nu\sigma}$ has been taken into account in (1). A set of three Nakanishi-Lautrup type auxiliary fields 
(${\cal B}, \, B, \,  B_\mu$) have been invoked to linearize the kinetic term and the 
gauge-fixing terms for the gauge fields $A_\mu$ and $B_{\mu\nu}$, respectively.  In the above, the 2-form:
$B^{(2)} = \frac{1}{2}\, B_{\mu\nu}\, (d x^\mu \wedge d x^\nu)$ defines the antisymmetric 
($B_{\mu\nu} = -\, B_{\nu\mu}$) tensor gauge field $B_{\mu\nu}$ and the 3-form: 
$H^{(3)} = d\, B^{(2)} = \frac{1}{3 !}\, H_{\mu\nu\lambda}\, (d x^\mu \wedge d x^\nu \wedge d x^\lambda)$
defines the field-strength tensor $H_{\mu\nu\lambda} = \partial_\mu\, B_{\nu\lambda}  + \partial_\nu\, B_{\lambda\mu}
+ \partial_\lambda\, B_{\mu\nu}$ for the antisymmetric tensor gauge field. Here $d = \partial_\mu \, d x^\mu$
[with $d^2 = \frac{1}{2 !} \, (\partial_\mu\, \partial_\nu - \partial_\nu\, \partial_\mu) \, (d x^\mu \wedge d x^\nu) = 0$]
is the exterior derivative of differential geometry. The scalar field $\phi$ is present in the 
theory due to the stage-one  reducibility for our 2-form field $B_{\mu\nu}$. 
Similarly, for the Abelian 1-form theory, we have the field-strength tensor $F_{\mu\nu} 
= \partial_\mu\, A_\nu - \partial_\nu\, A_\mu$ which is derived from the 2-form 
$F^{(2)} = d\, A^{(1)} = \frac{1}{2 !} \, F_{\mu\nu}\, (d x^\mu \wedge d x^\nu)$  where the 1-form  $A^{(1)} = A_\mu\, d x^\mu $
defines the vector potential $A_\mu$ of our Abelian 1-form gauge theory..

In the BRST invariant Lagrangian density (1), we have the Lorentz vector 
fermionic (i.e. $C_\mu^2 = \bar C_\mu^2 = 0, \, C_\mu\, C_\nu + C_\nu \, C_\mu = 0, \, 
C_\mu \bar C_\nu + \bar C_\nu \, C_\mu = 0, $ etc.) (anti-)ghost fields ($\bar C_\mu)C_\mu$
with ghost numbers $(-1) +1$, respectively. On the other hand, the (anti-)ghost fields ($\bar\beta)\,\beta $
are the ghost-for-ghost fields which are bosonic (i.e. $\beta^2 \neq 0, \bar\beta^2 \neq 0$) in nature
and they carry the ghost numbers $(- 2) +2$, respectively. The fermionic (i.e. $C^2 = 0, \, \bar C^2 = 0, \, 
C\, \bar C + \bar C \, C = 0$) (anti-)ghost fields ($\bar C)\, C$  are endowed with ghost numbers $(-1)+1$,  respectively. 
These {\it latter} (anti-)ghost fields correspond to the Abelian 1-form gauge field (within the framework of BRST formalism). 
The auxiliary (anti-)ghost fields $(\rho)\lambda$  of our system
also carry the ghost numbers (- 1)+1, respectively, because we note that 
$\rho = -\, (1/2) \, (\partial \cdot \bar C)$  and $\lambda =  (1/2) \, (\partial \cdot  C)$. 
These (anti-)ghost fields are required to maintain the sacrosanct  property of unitarity  in our BRST-invariant theory
which is valid at any arbitrary order of perturbative  computation for {\it all} 
the physical processes that are allowed by our BRST-quantized theory.

The above Lagrangian density (1) respects the following infinitesimal, continuous and off-shell nilpotent 
($s_b^2 = 0$) BRST symmetry transformations
\begin{eqnarray}
&& s_{b} B_{\mu \nu}=-\left(\partial_{\mu} C_{\nu} - \partial_{\nu} C_{\mu}\right), \qquad s_{b} C_{\mu}= -\,\partial_{\mu} \beta, 
\nonumber\\
&&s_{b} \bar{C}_{\mu}= - \,B_{\mu},\qquad 
s_{b} A_{\mu}= \partial_{\mu} C, \;\qquad s_{b} \bar{C}= B, \qquad \;
\nonumber\\
&&s_{b} \bar{\beta} = - \,\rho, \qquad \quad \;s_{b} \phi = + \,\lambda, \qquad \quad s_{b} F_{\mu \nu}= 0,  \nonumber\\
&& s_{b}\left[ \rho, \lambda, C, \beta, B_{\mu}, B, \mathcal{B}, H_{\mu \nu \lambda}\right]=0,
\end{eqnarray}
because we observe  that: 
\begin{eqnarray}
s_{b} \mathcal{L}_{B} &=& - \,\partial_{\mu}\Big[\left(\partial^{\mu} C^{\nu}-\partial^{\nu} C^{\mu}\right) B_{\nu}
+ \lambda \,B^{\mu} \nonumber\\
&+& \rho\, \partial^{\mu} \beta + B\,\partial^{\mu} C\Big]. 
\end{eqnarray}
As a consequence, the action integral $S = \int d^3 x \, {\cal L}_B$  remains invariant (i.e. $s_b\, S = 0$)
because all the physical fields vanish off as $x \rightarrow \pm \infty$. 
The analogous to the Lagrangian density ${\cal L}_B$,  we have an anti-BRST invariant Lagrangian density
$({\cal L}_{\bar B})$ as 
\begin{eqnarray}
{\cal L}_{\bar B} &=& \mathcal{B} \left(\frac{1}{2} \varepsilon^{\mu \nu \sigma} \partial_{\mu} B_{\nu \sigma}\right)-\frac{\mathcal{B}^{2}}{2}
+ \bar B^{\mu} \left(\partial^{\nu} B_{\nu \mu} + \partial_{\mu} \phi \right)
\nonumber\\
&-& \frac{\bar B^{\mu} \bar B_{\mu}}{2} +\frac{B^{2}}{2}   - B (\partial \cdot A)-\frac{1}{4} F^{\mu \nu} F_{\mu \nu} \nonumber\\
&-& \partial_{\mu} \bar{C} \, \partial^{\mu} C
+ \partial_{\mu} \,  \bar{\beta}\, \partial^{\mu} \beta +\left(\partial_{\mu} \bar{C}_{\nu}
 -\partial_{\nu} \bar{C}_{\mu}\right)\left(\partial^{\mu} C^{\nu}\right)
\nonumber\\  
  &+&(\partial \cdot \bar{C}+\rho)\,  \lambda+(\partial \cdot C-\lambda)\, \rho, 
\end{eqnarray}
where $\bar B_\mu$ is a new Nakanishi-Lautrup type auxiliary field that has been invoked to 
linearize  the gauge-fixing term for the Abelian antisymmetric tensor field $B_{\mu\nu}$ where we have taken into account 
$\phi \rightarrow - \,\phi$ for the sake of generality in  the gauge-fixing term. 
The above Lagrangian density $({\cal L}_{\bar B})$ respects the following off-shell nilpotent 
($s_{ab}^2 = 0$) anti-BRST symmetry transformations
\begin{eqnarray}
&& s_{ab} B_{\mu \nu}=-\left(\partial_{\mu} \bar C_{\nu} - \partial_{\nu} \bar C_{\mu}\right), \qquad s_{ab} \bar C_{\mu}= -\,\partial_{\mu} \bar \beta, 
\nonumber\\
&&s_{ab} {C}_{\mu}= + \,\bar B_{\mu}, \qquad 
s_{ab} A_{\mu}= \partial_{\mu} \bar C, \;\quad s_{ab} C = -\, B,
\nonumber\\
&&s_{ab} {\beta} = - \,\lambda, \quad \qquad \;s_{ab} \phi = + \,\rho, \qquad \quad s_{ab} F_{\mu \nu}= 0,  \nonumber\\
&& s_{ab}\left[ \rho, \lambda, \bar C, \bar \beta, \bar B_{\mu}, B, \mathcal{B}, H_{\mu \nu \lambda}\right]=0,
\end{eqnarray}
because we observe that $({\cal L}_{\bar B})$ transforms to a total spacetime derivative as follows: 
\begin{eqnarray}
s_{ab} \mathcal{L}_{\bar B} &=& - \,\partial_{\mu} \Big[\left(\partial^{\mu} \bar C^{\nu}-\partial^{\nu} \bar C^{\mu}\right) \bar B_{\nu}
-\, \rho \,\bar B^{\mu} 
\nonumber\\
&+& \lambda\, \partial^{\mu} \bar \beta + B\,\partial^{\mu} \bar C\Big]. 
\end{eqnarray} 
As a consequence, the action integral $S = \int d^3 x \, {\cal L}_{\bar B}$  remains invariant (i.e. $s_{ab}\, S = 0$) due to the 
Gauss divergence theorem because all the physical fields of our theory vanish off as $x \rightarrow \pm \infty$. 
Thus, we note that ${\cal L}_{\bar B}$ respects the nilpotent anti-BRST transformations $s_{ab}$.

We end this section with the following remarks. First of all, 
we note that the field strength tensors $H_{\mu\nu\lambda}$ and $F_{\mu\nu}$ (owing 
their origin to the exterior derivative $d$ of differential geometry) remain invariant 
under the nilpotent (anti-)BRST symmetry transformations. Second, the nilpotent versions of the (anti-)BRST symmetry transformations are valid in any 
arbitrary dimension of spacetime. Third, it is well-known that the Lagrangian density 
${\cal L}_{B}^{(1)} = -\, \frac{1}{4}\, F^{\mu\nu}\, F_{\mu\nu} + \frac{1}{2}\, B^2 - B\, (\partial \cdot A) - 
\partial_\mu \bar C \, \partial^\mu\, C$ for the Abelian 1-form gauge theory is invariant under the 
(anti-)BRST symmetry transformations which are off-shell nilpotent and absolutely anticommuting  in nature\footnote{The infinitesimal, continuous and off-shell  nilpotent (anti-)BRST symmetry transformations: $s_{ab} A_\mu = 
\partial_\mu \bar C, \;s_{ab} \bar C = 0, \; s_{ab} C = -\, B, \; s_{ab} B = 0$
and $s_{b} A_\mu = \partial_\mu  C, \;s_{b}  C = 0, \; s_{b} \bar C = +\, B, \; s_{b} B = 0$ satisfy the absolute anticommutativity
property (i.e. $\{ s_b, \; s_{ab} \} = 0$) without any recourse to the CF-type restriction because the {\it latter} is trivial in this case.}
without invoking any kind of CF-type restriction. On the other hand, the (anti-)BRST symmetry transformations for the 
Abelian 2-form theory are absolutely anticommuting only when one invokes the validity\footnote{It can be
readily checked that $\{s_b, \; s_{ab} \} \, B_{\mu\nu} = \partial_\mu (B_\nu - \bar B_\nu)  - \partial_\nu (B_\mu - \bar B_\mu)$
by taking into account the (anti-)BRST symmetry transformations in (6) and (3), respectively. The absolute anticommutativity property
(i.e. $\{ s_b, \; s_{ab} \} = 0$) is satisfied if and only if we utilize the expression for the CF-type restriction.} 
of the CF-type restriction: 
$B_\mu - \bar B_\mu + 2\, \partial_\mu\, \phi = 0$ which emerges from the EL-EoMs w.r.t. the auxiliary fields $B_\mu$ and 
$\bar B_\mu$ from the Lagrangian densities ${\cal L}_{ B}$ and ${\cal L}_{\bar B}$, respectively (see, e.g. [19]). 
Finally, we  would like to mention that the there are (anti-)co-BRST symmetry
transformations for the free Abelian 1-form theory in the two (1 + 1)-dimensions of spacetime (see, e.g. [20]). 
We have also shown that for the Abelian 2-form theory, there is existence of the nilpotent
(anti-)co-BRST transformations in the four (3 + 1)-dimensions of the spacetime (see, e.g. [10]).
We shall see, in our next section, that there is existence of the (anti-)co-BRST symmetry transformations for 
our present system (which is a combination of the free Abelian 1-form and 2-form theories) in the {\it three} 
(2 + 1)-dimensions of Minkowskian flat spacetime. \\


\section{(Anti-)co-BRST Symmetries: Salient Features}

We point out that the gauge-fixing terms ($\partial \cdot A$) and ($\partial^\nu\, B_{\nu\mu}$) for the 
Abelian 1-form and 2-form gauge fields owe their origin to the co-exterior derivative 
$\delta = \mp\, * \, d \, *$  (with $\delta^2 = 0$)
where $*$ is the Hodge duality operator on a given spacetime manifold. It is straightforward to note that 
$\delta \, A^{(1)} = +\, * \, d \, * \, A^{(1)} = (\partial \cdot A)$ and 
$\delta \, B^{(2)} = -\, * \, d \, * \, B^{(2)} = (\partial^\nu  B_{\nu\mu})\; d\, x^\mu$ 
because the operation of the co-exterior derivative of differential geometry on a given form 
reduces the degree of the form by {\it one}. We demand that, under the (anti-)co-BRST [or (anti-)dual-BRST]
symmetry transformations [$s_{(a)d}$], the total gauge-fixing term  
($\partial \cdot A$) and ($\partial^\nu\, B_{\nu\mu} \mp \partial_\mu \, \phi$)
should remain invariant. Against the backdrop of the above discussion, it can be checked that the under the following 
infinitesimal, continuous and  off-shell nilpotent (anti-)co-BRST
symmetry transformations  [$s_{(a)d}$], namely;
\begin{eqnarray}
&&s_{ad}\, B_{\mu\nu} = -\, \varepsilon_{\mu\nu\sigma} \, \partial^\sigma\, C, \qquad
s_{ad}\, C_\mu =\, \partial_\mu\, \beta, \nonumber\\
&&s_{ad}\, \bar C = {\cal B},  \quad s_{ad} \, \bar \beta =\, \rho, \quad s_{ad}\, \Big[\rho ,\, \lambda, \, C, \, \beta, \, \bar C_\mu, \nonumber\\
&&  (\partial^\nu\, B_{\nu\mu}), \, 
\phi, \, A_\mu, \, F_{\mu\nu},  B, \, {\cal B}, \,B_{\mu}, \, \bar B_{\mu} \Big] = 0, 
\end{eqnarray}
\begin{eqnarray}
&&s_{d}\, B_{\mu\nu} = -\, \varepsilon_{\mu\nu\sigma} \, \partial^\sigma\, \bar C, \qquad
s_{d}\, \bar C_\mu = -\, \partial_\mu\, \bar \beta, \nonumber\\
&&s_{d}\, C = -\, {\cal B},  \quad
s_{d}\, \beta = -\, \lambda, \quad s_{d}\, \Big[\rho ,\, \lambda, \, B, \,{\cal B},\, B_{\mu}, \, \bar B_{\mu}, \, \bar C,  \nonumber\\
 &&   
 \bar\beta, \,C_\mu, \,  A_{\mu}, \, (\partial^\nu\, B_{\nu\mu}), 
  F_{\mu\nu}, \, \phi \Big] = 0, 
\end{eqnarray}
{\it both} the Lagrangian densities $ {\cal L}_{B}$ and $ {\cal L}_{\bar B}$ transform in a similar fashion\footnote{It is pertinent
to point out that there is a non-trivial choice of the (anti-)co-BRST symmetry transformations for the vector field $A_\mu$ under which
the gauge-fixing term $(\partial \cdot A)$ remains invariant. These are: $s_{ad} A_\mu = \pm\, \varepsilon_{\mu\nu\sigma} \, \partial^\nu C^\sigma, \;
 s_{d} A_\mu = \pm\, \varepsilon_{\mu\nu\sigma} \, \partial^\nu \bar C^\sigma$. However, this kind of choice does {\it not} lead to the
 {\it overall}  symmetry of the Lagrangian densities $ {\cal L}_{B}$ and $ {\cal L}_{\bar B}$. Moreover, we find that the kinetic term 
(i.e. $-\, \frac{1}{4}\,F_{\mu\nu}\, F^{\mu\nu}$) for the vector field $A_\mu$ transforms to an expression that contains 
{\it three} derivatives (which is pathological).}. To be precise, 
we note that  $ {\cal L}_{B}$ and $ {\cal L}_{\bar B}$ transform under the off-shell nilpotent (anti-)co-BRST symmetry transformations 
$[s_{(a)d}]$ to the total spacetime derivatives as follows:
\begin{eqnarray}
&&s_d\, {\cal L}_{B} = -\, \partial_\mu \, [{\cal B}\, \partial^\mu\, \bar C + \lambda \, \partial^\mu\, \bar\beta] 
\equiv    
s_d \,  {\cal L}_{\bar B}, \nonumber\\
&&s_{ad}\, {\cal L}_{B} = -\, \partial_\mu \, [{\cal B}\, \partial^\mu\, C - \rho \, \partial^\mu\, \beta] 
\equiv    s_{ad}\,  {\cal L}_{\bar B}.
\end{eqnarray}
The above observation establish that the action integrals 
$S_1 = \int d^3 x \, {\cal L}_{\bar B}$ and $S_2 = \int d^3 x \, {\cal L}_{\bar B}$ {\it both} remain invariant under the 
(anti-)dual-BRST symmetry transformations  (i.e. $s_{(a)d}\, S_1 = s_{(a)d}\, S_2 = 0$). This happens due to the Gauss 
divergence theorem because all the physically well-defined  fields vanish off as $x \rightarrow \pm \infty$.

There are a few interesting observations when we focus on the 
infinitesimal, continuous and off-shell nilpotent (anti-)BRST symmetry 
transformations [cf. Eqs. (6), (3)] and the (anti-)co-BRST symmetry
transformations [cf. Eqs. (9), (8)] in their operator forms. For instance, we note that the following 
anticommutators are true, namely; 
\begin{eqnarray}
\{s_d, \, s_{ad} \} = 0, \qquad  \qquad  \{s_d, \, s_{ab} \} = 0, 
\nonumber\\
\{s_b, \, s_{ad} \} = 0, \qquad \qquad \{s_d, \, s_{ab} \} = 0. 
\end{eqnarray}
All the above anticommutators are automatically satisfied {\it but} for the validity of 
$\{s_b, \, s_{ab} \} = 0$. In fact, the {\it latter} anticommutator (i.e. $\{s_b, \, s_{ab} \} = 0$) is {\it also} 
automatically satisfied for {\it all} the fields of the theory except the antisymmetric  tensor  gauge field $B_{\mu\nu}$. 
In other words, the anticommutator: $\{s_b, \, s_{ab} \}\, B_{\mu\nu} = 0$ is satisfied if and only if the CF-type 
restriction: $B_\mu -\, \bar B_\mu + 2\, \partial_\mu\, \phi = 0$ is invoked for its validity. 
Ultimately, we note that out of the {\it four}  nilpotent (anti-)BRST and (anti-)co-BRST symmetry transformations of our theory, only the 
following {\it two} non-trivial anticommutators, namely; 
\begin{eqnarray}
\{s_b, \, s_{d} \} \neq 0, \qquad \qquad \{s_{ad}, \, s_{ab} \} \neq 0,
\end{eqnarray}
are the ones that lead to the definitions of the bosonic symmetry transformations.

We wrap-up this section with the following concluding remarks. First of all, we note that the gauge-fixing terms 
[owing their origin to the co-exterior (or dual-exterior) derivative of differential geometry] remain invariant
under the (anti-)co-BRST symmetry transformations\footnote{This is precisely the reason behind calling these off-shell
nilpotent symmetries as the (anti-)co-BRST [or the (anti-)dual-BRST] symmetry transformations because, under the (anti-)BRST
symmetry transformations, the field-strength tensors for the Abelian 1-form and 2-form gauge fields $F_{\mu\nu}$ 
and $H_{\mu\nu\sigma}$ (owing their origin to the
exterior derivative) remain invariant.}. Second, the absolute anticommutativity property
(i.e. $\{s_d, \, s_{ad} \} = 0$) is automatically satisfied without invoking any kind of CF-type restriction. 
Third, both the coupled Lagrangian densities respect both (i.e. co-BRST and anti-co-BRST) symmetry transformations.
Fourth, it is interesting to point out that the variation of the kinetic term for the Abelian 2-form field under the 
(anti-)co-BRST symmetry transformations is compensated by the variation of the ghost term of the Abelian 
1-form theory. 
This observation should be contrasted against the (anti-)BRST  symmetry 
transformations, under which, the Lagrangian densities for the 1-form
and 2-form theories transform, on their {\it own}, to the total spacetime 
derivatives. Finally, only {\it two} of the anticommutators between the (anti-)BRST and (anti-)co-BRST symmetry transformations are 
non-trivial [cf. Eq. (12)]. \\


\section{Bosonic Symmetry Transformations: Uniqueness}

We have noted that  there are two non-trivial anticommutators [cf. Eq. (12)] between the off-shell 
nilpotent (anti-)BRST and (anti-)co-BRST symmetry transformations which define the following infinitesimal and continuous  bosonic symmetry transformations: 
\begin{eqnarray}
s_w = \{s_b, \, s_d \}, \qquad \qquad s_{\bar w} = \{ s_{ad}, \, s_{ab} \}. 
\end{eqnarray}
We shall see that only one of the above bosonic symmetry transformations is the independent bosonic symmetry of our theory
which turns out to be  {\it unique}. To corroborate this statement, we note that the following infinitesimal and continuous
bosonic symmetry transformations (i.e. $s_w = \{s_b, \, s_d \}$) are true, namely; 
\begin{eqnarray}
&&s_w \, B_{\mu\nu}  = -\, \varepsilon_{\mu\nu\sigma}\, \partial^\sigma\, B, \qquad s_w \, \bar C_\mu = \partial_\mu\, \rho, \nonumber\\
&&s_w\, C_\mu = \partial_\mu\, \lambda,  \qquad \qquad 
s_w\, A_\mu = -\, \partial_\mu \, {\cal B}, \nonumber\\ 
&&s_w \, [\rho, \, \lambda, \, B, \, B_\mu, \bar B_\mu, \, {\cal B}, \, 
C,\, \bar C, \, \beta, \, \bar\beta, \, \phi] = 0,
\end{eqnarray}
where we have used the symmetry transformations that have been listed in (3) and (9) to determine the transformation (14)
which corresponds to $s_w = \{s_b, \, s_d\}$. On the other hand, we obtain the following symmetry transformations for the fields of 
our theory (corresponding to the bosonic symmetry transformations: $s_{\bar w} = \{s_{ab}, \, s_{ad}\}$): 
\begin{eqnarray}
&&s_{\bar w} \, B_{\mu\nu}  =  \varepsilon_{\mu\nu\sigma}\, \partial^\sigma\, B, \qquad
 s_{\bar w} \, \bar C_\mu =  -\, \partial_\mu\, \rho, \nonumber\\
&&s_{\bar w}\, C_\mu = -\, \partial_\mu\, \lambda,\qquad \qquad 
s_{\bar w}\, A_\mu = \partial_\mu \, {\cal B}, \nonumber\\ 
&&s_{\bar w} \, [\rho, \, \lambda, \, B, \, B_\mu, \bar B_\mu, \, {\cal B}, \, 
C,\, \bar C, \, \beta, \, \bar\beta, \, \phi] = 0.
\end{eqnarray}
A close and careful look at (14) and (15) shows that we have the validity of $s_w + s_{\bar w} = 0$. In other words, we 
have a {\it unique} bosonic symmetry in the theory. This claim is corroborated, once again, in the following observations: 
\begin{eqnarray}
s_w \,{\cal L}_{B} = \partial_\mu\, \Big [B\, \partial^\mu\, {\cal B} - 
{\cal B} \, \partial^\mu\, B - \rho\, \partial^\mu\, \lambda\, + (\partial^\mu\, \rho)\, \lambda \Big], \nonumber\\
s_{\bar w} \,\cal{L}_{B} = -\,  \partial_\mu\, \Big [B\, \partial^\mu\, {\cal B} - 
{\cal B} \, \partial^\mu\, B - \rho\, \partial^\mu\, \lambda\, + (\partial^\mu\, \rho)\, \lambda \Big], 
\end{eqnarray}
which establish that  $(s_w + s_{\bar w})\, {\cal L}_{B} = 0$. In other words the operator equation: 
$s_w + s_{\bar w} = 0$ implies that only {\it one} bosonic symmetry 
transformation is independent. This proves the {\it uniqueness} of 
the local, infinitesimal and continuous bosonic symmetry transformations
of our 3D system of a combination of the free Abelian 1-form and 2-form gauge theories.

We end this section with the following concluding remarks. First of all, we note that, 
under the bosonic symmetry transformations (14) and/or (15), all the ghost fields either
do {\it not} transform at all {\it or} transform up to a U(1) gauge symmetry-type of transformations. 
The {\it latter} observation is true  {\it only} for the Lorentz vector (anti-)ghost fields $(\bar C_\mu)\, C_\mu$. 
Second, out of the {\it four} off-shell nilpotent symmetries, only {\it two} anticommutators 
(out of {\it six} possible anticommutators [cf. Eqs. (11), (12)]) are non-trivial [cf. Eq. (12)] which define the 
bosonic symmetry in our theory. Out of these two bosonic symmetry transformations, only {\it one} of 
them is independent (e.g. $s_w$) which proves its uniqueness. Finally, a close look at (11), (12) and (13) 
establishes  that the following commutators are true, namely; 
\begin{eqnarray}
&&[s_w, \, s_b] = 0, \qquad [s_w, \, s_{ab}] = 0, \nonumber\\
&&[s_w, \, s_d] = 0, \qquad [s_w, \, s_{ad}] = 0, 
\end{eqnarray}
where we have taken into account  the off-shell nilpotency $\big(s_{(a)b}^2 = 0, \; s_{(a)d}^2 = 0\big)$ property of the 
(anti-)BRST and (anti-)co-BRST symmetry transformations.\\


\section{Ghost-Scale Symmetry Transformations: Key Observations and Their Implications}

It is straightforward to check that under the following ghost-scale transformations 
\begin{eqnarray}
&&\beta \longrightarrow e^{+ 2\, \Sigma} \, \beta,  \qquad \qquad
\bar \beta \longrightarrow e^{-\,  2\, \Sigma} \,\bar\beta, \nonumber\\
&&C_\mu \longrightarrow e^{+ \Sigma} \, C_\mu,  \qquad \quad
\bar C_\mu \longrightarrow e^{-\, \Sigma} \,  \bar C_\mu, \nonumber\\
&& C \longrightarrow e^{+ \Sigma} \,  C, \qquad \qquad \; \;
\bar C \longrightarrow e^{-\,  \Sigma} \,  \bar C, \nonumber\\
&&\lambda \longrightarrow e^{ + \Sigma} \,  \lambda, \qquad \qquad\quad
\rho \longrightarrow e^{-\, \Sigma} \,  \rho, \nonumber\\
&&\Phi \longrightarrow e^0\, \Phi \, (\Phi = 
A_\mu, \, B_{\mu\nu}, \, \phi, \, \bar B_\mu, \, B_\mu, \, {\cal B}, \,B),  
\end{eqnarray}
where $\Sigma$ is a global (i.e. spacetime independent) scale transformation parameter, the Lagrangian densities
${\cal L}_{B}$ and ${\cal L}_{\bar B}$ remain invariant. For the sake of brevity, 
if we choose the scale parameter $\Sigma = 1$, the infinitesimal version $(s_g)$ of the {\it above} ghost-scale 
symmetry transformations reduce to the following simple forms, namely;
\begin{eqnarray}
&&s_g\, \beta = + 2\, \beta, \quad s_g\, \bar \beta = -\, 2\, \bar\beta, \quad 
s_g\, C_\mu = +\, C_\mu, \nonumber\\
&& s_g\, \bar C_\mu = -\, \bar C_\mu,  \quad
s_g\, C = + \, C, \quad s_g\, \bar C = -\, \bar C, \nonumber\\
&&s_g\, \lambda =  \lambda, \quad 
s_g\, \rho = -\, \rho, \quad s_g\, \Phi = 0.
\end{eqnarray}
It is straightforward to check that, under the above infinitesimal ghost scale symmetry transformations, we obtain: 
$s_g\, {\cal L}_{B} = 0, \; s_g\, {\cal L}_{\bar B} = 0$. Hence, the action integrals corresponding to the 
Lagrangian densities ${\cal L}_{B}$ and ${\cal L}_{\bar B}$ 
{\it also} remain invariant under $s_g$.

We end this section with a couple of clinching remarks. First of all, we note that only the 
(anti-)ghost basic and auxiliary [i.e. $\lambda =  \frac{1}{2} \, (\partial \cdot  C)$,
 $\rho = -\, \frac{1}{2} \, (\partial \cdot \bar C)$] fields transform under the ghost-scale 
 symmetry transformations and rest of the basic fields (e.g. $\Phi = B_{\mu\nu}, \, 
 A_\mu, \,\phi,\, B, \,\, {\cal B},\, B_\mu,  \; \bar B_\mu$) of the theory do {\it not} transform at all. 
 Second, we note that the operator form of the infinitesimal  symmetries obey the following very useful relationships: 
\begin{eqnarray}
&&[s_g, \, s_w] = 0, \quad [s_g, \, s_b] = + \,s_b, \quad [s_g, \, s_d] = -\,  s_d, \nonumber\\
&&[s_g, \, s_{ab}] = -\, s_{ab},  \qquad  [s_g, \, s_{ad}] = +\, s_{ad}. 
\end{eqnarray}
The above commutators establish the fact that the pairs ($s_b, \, s_{ad}$) and ($s_{ab}, \, s_d$) obey exactly similar kinds of 
algebra w.r.t. $s_g$ {\it and} $s_w$ commutes with $s_g$. Thus, we note that the bosonic symmetry transformation $s_w$
commutes with {\it all} the symmetry operators of our theory.\\


\section{Algebraic Structures: Cohomological Operators}

In this section, we provide the physical realizations of the de Rham cohomological operators of the 
differential geometry at the algebraic level and establish that there is two-to-one mapping 
between the continuous symmetry operators of our theory and the cohomological operators. 
The {\it latter} are defined on a compact spacetime manifold without a boundary and they form a 
set of three operators ($d, \, \delta, \Delta$) where $d = \partial_\mu\, d\, x^\mu$
[with $d^2 = \frac{1}{2 !}\, (\partial_\mu\, \partial_\nu - \partial_\nu\, \partial_\mu)\, (d\, x^\mu \wedge d\, x^\nu) = 0$]
is the exterior derivative, $\delta = \pm \, *\, d \, *$ (with $\delta^2 = 0$) is the 
co-exterior derivative and $\Delta = (d + \delta)^2 = \{d, \, \delta\}$ is the 
Laplacian operator. They obey an algebra which is properly known as the Hodge algebra of 
differential geometry. This algebra is: 
\begin{eqnarray}
&&d^2 = 0, \qquad \delta^2  = 0,  \qquad \Delta = (d + \delta)^2 = \{d, \, \delta\}, \nonumber\\
&& [\Delta, \, d] = 0, \qquad [\Delta, \, \delta] = 0, \qquad \{d, \, \delta\} \neq 0.
\end{eqnarray} 
Thus, we note that the Laplacian operator commute with both the nilpotent ($d^2 = \delta^2 = 0$) exterior and 
co-exterior derivatives of the differential geometry.

At this juncture, we collect all the algebraic structures 
of the symmetry operators which we have discussed in our 
previous sections. In their full blaze of glory, the algebraic structures of the symmetry operators are as follows: 
\begin{eqnarray}
&&s_b^2 = 0, \quad s_d^2 = 0, \quad s_{ab}^2 = 0, \quad s_{ad}^2 = 0,
\nonumber\\ 
&&s_w = \{s_b, \, s_d\} = -\, \{s_{ad}, \, s_{ab}\}, \nonumber\\
&&[s_w, \, s_r] = 0, \quad r = b, \, ab, \, d, \, ad, \, g, \, w, 
\nonumber\\
&& \{s_b, \, s_{ad}\} = 0, \qquad  \{s_b, \, s_{ab}\} = 0, \nonumber\\
&&  \{s_{ab}, \, s_{d}\} = 0, \qquad \{s_d, \, s_{ad}\} = 0, \nonumber\\
&& [s_g, \, s_b] = + s_b, \qquad [s_g, \, s_{ab}] = -\, s_{ab},\nonumber\\
&& [s_g, \, s_{ad}] = + s_{ad},  \quad [s_g, \, s_w] = 0,\nonumber\\
&&[s_g, \, s_d] = -\,  s_d. 
\end{eqnarray}
A comparison between the equations  (21) and (22) establishes the two-to-one mapping between the symmetry operators of (22) and the 
cohomological operators (21) as 
\begin{eqnarray}
(s_b, \, s_{ad}) \longrightarrow d, \quad (s_d, \, s_{ab}) \longrightarrow \delta,  \quad
(s_w, \, s_{\bar w})  \longrightarrow \Delta, 
\end{eqnarray}
where, as has been established in section four, the bosonic symmetry operator $s_{\bar w} = -\, s_{w}$.

We conclude this section with the remarks that the algebraic relationships between the infinitesimal 
and continuous ghost-scale symmetry operator ($s_g$) and other infinitesimal continuous symmetry
operator (e.g. $s_b,\, s_{ab}, \, s_d, \, s_{ad}, \, s_w$) in equation (22) are important because they 
encompass in their folds the  fact that symmetry operators ($s_b, s_{ad}$) operating on a specific field
raise the ghost number by one. On the other hand, the symmetry operators ($s_d, \, s_{ab}$) lower the 
ghost number by one when they operate on a specific field. 
In contrast to the above observations, the unique bosonic symmetry operator ($s_w$) does not change the ghost number of a 
specific field when it operates on that field. In terms of the Noether conserved charges, corresponding to the 
above infinitesimal continuous symmetries, the above observations are translated into the quantum Hilbert 
space of states which we plan to do in our future endeavor.  \\


\section{Conclusions}

In our present investigation, we have concentrated on the continuous symmetry transformations for the coupled (but equivalent)
Lagrangian densities of the {\it combined} system of the free 3D Abelian 1-form and 
2-form gauge theories within the framework of BRST formalism and established that the 
algebra, obeyed by the symmetry transformation operators, is reminiscent of the Hodge algebra 
that is respected by the de Rham cohomological operators of the differential geometry. We have also
demonstrated that the continuous transformations  of our theory correspond to the cohomological 
operators ($d, \, \delta, \Delta$) at the algebraic level and there is two-to-one mapping
between them (cf. Eq. (26)]. Despite our sincere efforts, we have {\it not} yet been able to show the existence 
of the discrete symmetries in our theory to establish the physical realizations of the Hodge duality $*$ operation
which connects the (anti-)co-BRST and (anti-)BRST symmetry transformations (see, e.g. [8-10],[20]). In other words, we
have not been able to provide the physical realization(s) of the connection between the 
(co-)exterior derivatives: $\delta = \pm\, *\, d\, * $ 
of the differential geometry in the language of symmetry operators of our theory.

We have reasons to believe that our present {\it odd} dimensional (i.e. $D = 3$) example for Hodge theory might {\it not}
be endowed with the discrete duality symmetry transformations because there is {\it no} existence of a pseudo-scalar field
in our {\it present} theory which is dual to the scalar field $\phi$ that exists in the gauge-fixing term for the
Abelian 2-form field [cf. Eqs. (1),(5)]. This is {\it not} the case with the 2D St${\ddot u}$ckelberg-modified Proca theory (see, e.g. [20] for details)
as well as the 4D St${\ddot u}$ckelberg-modified massive as well as massless Abelian 2-form theories (see, e.g. [8-10] for details)
which are a set of tractable field-theoretic models of Hodge theory. To be precise, we have the existence of 
(i) the scalar and pseudo-scalar fields in
the context of the 2D modified Proca theory [20],  and  (ii) the pair scalar and pseudo-scalar as well as 
the pair vector and axial-vector in the context of the 4D Abelian 2-form {\it modified} massive gauge theory (see, e.g. [10]).
This is why there is existence of a set of duality symmetry transformations in these {\it even} dimensional models of Hodge theory.

We would like to lay emphasis on the fact that in all the {\it even} dimensional examples for Hodge theory [8-10,20], there is
existence of some bosonic fields with negative kinetic terms which obey the usual Klein-Gordon equation of motion. The {\it latter}
observation demonstrates  that they are endowed with the well-defines rest masses. Such kinds of ``exotic'' fields have become quite popular
in the realm of THEP as well as in the domain of cosmological models of the Universe. For instance, the fields with
negative kinetic terms with or without mass are (i) a possible  set of candidates for dark matter and dark energy 
(see. e.g. [21,22] for details), and (ii) a set of fields that has been christened  as the ``phantom'' or ``ghost''
fields in the contexts of the cyclic, bouncing and self-accelerated cosmological models of the Universe (see, e.g. [23-25])
which are responsible for the modern observation of the accelerated expansion of the Universe. It is interesting 
to point out that, for our present system of an odd-dimensional 
field-theoretic example for Hodge theory, there is {\it no} room for the existence of such kinds of fields with negative kinetic terms.
Same is the case with our earlier works on the 1D toy models and 
${\mathcal N}  = 2$ SUSY quantum mechanical systems that are examples
for Hodge theory [11, 12. 14, 15].

We have, in our present endeavor,
{\it not} computed the conserved Noether charges and {\it not} studied the algebra followed by them. In addition, 
we have {\it not} yet been able to provide the physical realization(s) of the Hodge duality $*$ operation of
differential geometry\footnote{ In our earlier works (see, e.g. [8-12]), we have provided the physical
realization(s) of the algebraic relationships: $\delta = \pm\, *\, d\, * $ in terms of the nilpotent (anti-)co-BRST and (anti-)BRST 
transformations where the $*$ has corresponded to a couple of useful  discrete symmetry transformations in our theories. For our present 3D system,
we have {\it not} yet been able to obtain the discrete symmetry transformations. Perhaps, the latter do {\it not} exist for our 3D theory.}. 
Thus, at our current level of understanding, our present 3D system
should be treated as the field-theoretic example for {\it quasi}-Hodge theory.
In our future investigation [26], we plan to concentrate on these issues and establish, from all 
aspects of the theoretical angles,  that our present {\it combined} system of the free 3D Abelian 1-form 
and 2-form theories is indeed a perfect field-theoretic example for Hodge theory in the odd (i.e. $D = 3$)
dimension of spacetime. If this turns out to be true, it will be a completely {\it new} result as far as our studies on the 
BRST approach to the existence of the tractable field-theoretic models of Hodge theory are concerned.  \\

\vskip 0.5cm        

\noindent
{\bf Acknowledgments}\\

\noindent
One of us (RPM)  learnt the basics of the special theory of relativity and elementary particle physics from Prof. C. P. Singh who was
an excellent teacher.
Both the authors, very humbly and respectfully,  dedicate their present work  to the memory of Prof. C. P. Singh who was
responsible for the revival of the Nuclear and  High Energy Physics Section at the Physics Department of the
Banaras Hindu University
and who passed away in the recent past. Useful comments by our esteemed Reviewer are thankfully acknowledged, too.\\

\vskip 0.5 cm      
\noindent
{\bf Conflicts of Interest}\\
\vskip 0.1cm
\noindent
The authors declare that there are no conflicts of interest. \\

\vskip 0.5 cm
\noindent
{\bf Data Availability}\\
\vskip 0.1cm
\noindent
No data were used to support this study.\\

\end{document}